%% file: Main.tex
\documentclass{INTERSPEECH2023}
\usepackage{cite}
\usepackage{setspace}
\usepackage{multirow}
\usepackage{multicol}
\usepackage{algorithm}
\usepackage{enumitem}
\usepackage{subcaption}
\usepackage{nicematrix}
\usepackage{booktabs}       
\interspeechcameraready

\title{Pruning Self-Attention for Zero-Shot Multi-Speaker Text-to-Speech}
\name{Hyungchan Yoon$^1$, Changhwan Kim$^1$, Eunwoo Song$^2$, Hyun-Wook Yoon$^2$, Hong-Goo Kang$^1$}
\address{
  $^1$Department of Electrical and Electronic Engineering, Yonsei University, Seoul, Korea,
  \\$^2$NAVER Cloud, South Korea}
\email{[hcy71, chkim]@dsp.yonsei.ac.kr, [eunwoo.song, hyunwook.yoon]@navercorp.com, hgkang@yonsei.ac.kr}

\begin{document}

\maketitle
 
\input{0_Abstract.tex}
\vspace{-3pt}
\input{1_Introduction.tex}
\vspace{-3pt}
\input{2_Related.tex}
\vspace{-3pt}
\input{3_Method.tex}
\vspace{-3pt}
\input{4_Experiment.tex}
\vspace{-4pt}
\input{5_Conclusion.tex}
\vspace{-4pt}
\input{6_Ack.tex}

\bibliographystyle{IEEEtran}
\bibliography{mybib}

\end{document}

%% file: 0_Abstract.tex
\begin{abstract}
\vspace{-2pt}
For personalized speech generation, a neural text-to-speech (TTS) model must be successfully implemented with limited data from a target speaker. 
To this end, the baseline TTS model needs to be amply generalized to out-of-domain data (i.e., target speaker's speech).
However, approaches to address this out-of-domain generalization problem in TTS have yet to be thoroughly studied.
In this work, we propose an effective pruning method for a transformer known as \emph{sparse attention}, to improve the TTS model's generalization abilities.
In particular, we prune off redundant connections from self-attention layers whose attention weights are below the threshold.
To flexibly determine the pruning strength for searching optimal degree of generalization, we also propose a new differentiable pruning method that allows the model to automatically learn the thresholds. 
Evaluations on zero-shot multi-speaker TTS verify the effectiveness of our method in terms of voice quality and speaker similarity.
\end{abstract}
\vspace{-4pt}

%% file: 1_Introduction.tex
\section{Introduction}
\vspace{-1pt}
With the advancement of deep learning technologies, recent studies in text-to-speech (TTS) have shown a rapid progress. In terms of generation quality, single- and multi-speaker TTS models can synthesize human-like voices with sufficient training data from the target speaker(s)\cite{shen2018natural,ren2020fastspeech,multi2020, kim2020glow, kim2021conditional}. 
Further, several few- or zero-shot multi-speaker TTS models have recently been developed to synthesize out-of-domain (OOD) speech with limited data from the target speaker\cite{chen2020adaspeech, min2021meta, zhao2022nnspeech, wu2022adaspeech, kim2022transfer, huang2022generspeech}. 
These models are trained using a large multi-speaker dataset to learn a general TTS mapping relationship conditioned on speaker representations. Then, they are either additionally fine-tuned with a few samples of the target speaker (few-shot) or used directly (zero-shot) for synthesis. 

Especially, zero-shot multi-speaker TTS models\cite{zhao2022nnspeech,wu2022adaspeech,kim2022transfer,huang2022generspeech} are widely being studied due to their unique advantage of not requiring any training data from the target speaker. A common approach of these models is to extract the speaker representations from reference speech using a reference encoder\cite{min2021meta,jia2018transfer,chien2021investigating}. These representations contain various prosodic characteristics
such as pronunciation style, speed\cite{wang2018style,skerry2018towards} of the reference
speech, as well as speaker identity. As such, the speaker representation is learned to play a crucial role as a latent vector that determines the prosodic characteristics of the synthesized speech during training. During inference, the speaker representation is extracted from the voice of the unseen speaker, enabling the generation of the desired voice.

However, zero-shot multi-speaker TTS models face the problem of domain mismatch between training and inference, unlike conventional TTS models that aim to synthesize only in-domain speech (i.e., speech from seen speakers).
Specifically, the latter must be generalized only to the unseen text, whereas the former must generalized not only to the unseen text but also to the reference speech of unseen speakers. Therefore, the challenge of improving synthesis performance in zero-shot multi-speaker TTS lies in generalizing the TTS models to OOD data, which refers to speech from unseen speakers.

One additional challenge faced by zero-shot multi-speaker TTS models is that they require varying levels of generalization ability depending on the dataset they are trained on. When there is a high degree of domain mismatch between the training and test data, such as differences in recording environments, the models require more generalization to prevent overfitting. Conversely, when there is little domain mismatch, over-generalization can lead to degraded performance. Therefore, finding the optimal strength of generalization is crucial for improving the synthesis performance of these models. However, current zero-shot multi-speaker TTS models lack a systematic approach to this problem and have difficulty controlling the generalization strength once developed. While adjusting the number of parameters is a classical approach to controlling generalization\cite{giles1994pruning}, it can be a manual and time-consuming process.

To this end, we propose a new controllable generalization method for zero-shot multi-speaker TTS models. In particular, we focus on the transformer\cite{vaswani2017attention}, which is the foundation for many TTS models.
Our method draws on previous studies in various research fields (such as image generation and speech recognition) demonstrating the effectiveness of optimizing the self-attention module in a generalization objective\cite{Child2019sparsetrans,Tay2020SparseSA,roy-etal-2021-efficient,kim2021generalizing,Kim2022LearnedTP}. 
In particular, they enhanced generalization abilities by adding sparsity to the self-attention connections.
For instance, Child et al.\cite{Child2019sparsetrans} factorized the self-attention matrix into sparse subsets, and Kim et al.\cite{kim2021generalizing} proposed removing the low-weight connections during inference.

In this study, we design a sparse attention method for zero-shot multi-speaker TTS to successfully solve its OOD generalization problem. The method is implemented by pruning off the connections from self-attention layer; we also propose a \emph{differentiable pruning} technique that can easily control the degree of generalization. 
Our contributions are outlined below: 
\begin{itemize}[leftmargin=*]
\setlength\itemsep{0.1em}
\item \textbf{New Application.} We apply the sparse attention mechanism to the TTS model, which eliminates redundant connections from the self-attention layer. Because the TTS model is trained under a condition that only uses high-weight residual connections, the sparse attention mechanism significantly improves its generalization ability. In particular, adding sparsity to the self-attention module reduces the number of parameters engaged in the overall TTS training by preventing backpropagation of gradients through low-weight connections, which alleviates overfitting. 
\item \textbf{Novel Pruning Technique.} We explore optimal pruning techniques for the sparse attention. We first introduce a vanilla pruning approach that eliminates the connections whose attention weights are below a predetermined threshold.
To flexibly adjust the pruning strength in case of various degrees of domain mismatch, we further propose a differentiable pruning method that adopts learnable thresholds.
\item \textbf{Performance.} Experiments on zero-shot TTS show that our proposed method notably improves the performance of OOD speech synthesis\footnote{Audio samples are available at: \normalfont{\url{https://hcy71o.github.io/SparseTTS-demo/}}}.
\end{itemize}

%% file: 2_Related.tex
\section{Related Works}
Owing to the increasing demand for customized voice synthesis, the OOD generalization problem has recently been studied in zero-shot multi-speaker TTS works.
StyleSpeech \cite{min2021meta} used meta-learning to make a TTS model effectively adapt to OOD voice and conditioned speaker representations to the model using few variables to minimize the domain mismatch.
For the same purpose, nnSpeech \cite{zhao2022nnspeech} introduced a speaker-guided conditional varational autoencoder to define speaker representations as Gaussian latent variables rather than high-dimensional embeddings.
Furthermore, GenerSpeech \cite{huang2022generspeech} leveraged wav2vec2.0 \cite{baevski2020wav2vec}, a contrastive model learned with numerous speech data, to obtain more robust speaker representations. 
Unlike the abovementioned approaches, we use the self-attention pruning method to directly generalize the basic architecture (i.e., transformer) of the TTS model, implying that it is applicable to other models with minimal modifications.

%% file: 3_Method.tex
\section{Proposed Method}
\label{Proposed Method}
\begin{figure}[t!]
    \vspace*{-5pt}
    \centerline{\includegraphics[width=.42\textwidth]{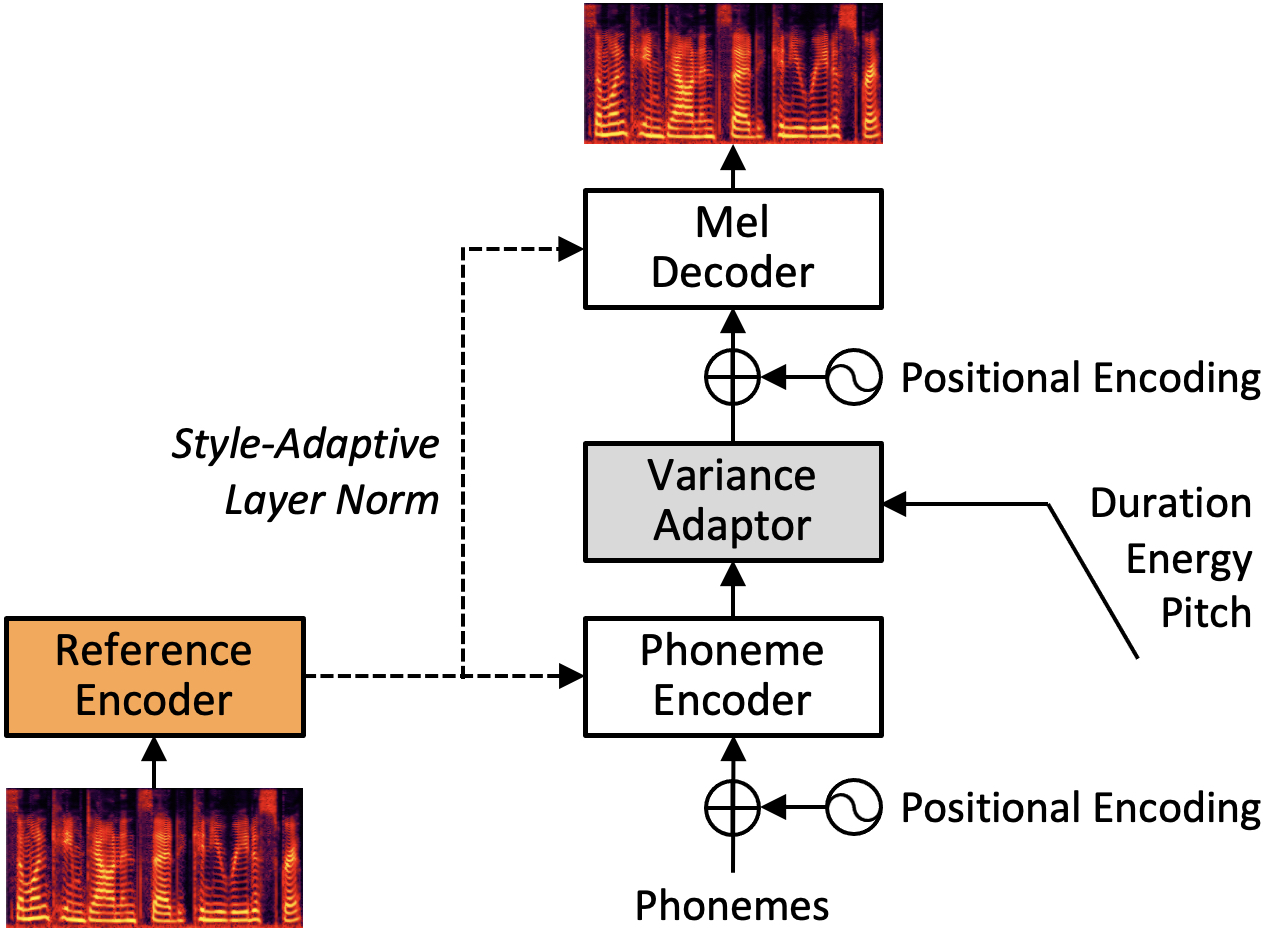}}
    \caption{\normalfont{Overview of StyleSpeech. The speaker representation is extracted from the reference encoder and provided to the encoder and decoder via Style-Adaptive Layer Normalization technique.}}
    \label{fig:overview}
\vspace{-8pt}
\end{figure}
We selected StyleSpeech\cite{min2021meta} as a baseline because it is a representative zero-shot multi-speaker TTS model built on a non-autoregressive transformer. As depicted in Fig \ref{fig:overview}, its architecture comprises a transformer-based phoneme encoder and mel-spectrogram decoder, a variance adaptor, and a reference encoder. The variance adaptor, located between the encoder and decoder, predicts the pitch, energy, and duration from phoneme-level embeddings; it then expands these embeddings to frame-level using the predicted duration values. The reference encoder extracts a speaker representation from the input reference speech and conditions it to the encoder and decoder via Style-Adaptive Layer Normalization\cite{min2021meta} technique.
More details, including loss terms and model configurations, are presented in\cite{min2021meta,ren2020fastspeech}.

\vspace{-2pt}
\subsection{Sparse Attention}
\vspace{-1pt}
We implement sparse attention by pruning redundant connections, and we only apply it to the decoder for the following two reasons: 1) The sequence length ($N$) of the decoder (frame-level) is much longer than that of the encoder (phoneme-level), indicating that the decoder has a significantly larger number of self-attention connections ($N\times N$) than the encoder; 
as a result, the decoder self-attention module requires more sparsity to be generalized. 2) According to our investigation, applying sparse attention to the encoder rather degrades the model performance because it reduces the modeling capacity of the original self-attention module.
We define sparse masks and apply them to all the attention heads of the decoder self-attention modules. 
Depending on the mask generation methods, we propose two types of pruning techniques: \textbf{vanilla} and \textbf{differentiable}.

\vspace{-3pt}
\subsubsection{Vanilla Pruning}
\vspace{-1pt}
Given queries $Q$ and keys $K$ obtained by two linear transformations $W_q$ and $W_k$, respectively, to the input sequence $\mathbf{x}$,
\begin{equation}
    {Q}= {W_q}\mathbf{x},\,{K}= {W_k}\mathbf{x},
\end{equation}
we first denote the attention probability of the $h$-th head of the multi-head self-attention layer \cite{vaswani2017attention} as $\mathcal{A}_{h}$:
\begin{equation}
\mathcal{A}_{h}(i,j)= softmax\left(\frac{{Q_h}{K_h}^T}{\sqrt{d}} \right)_{(i,j)},
\end{equation}
where $Q_h$ and $K_h$ are the queries and keys of the $h$-th head, respectively, and $d$ is their dimension. $\mathcal{A}_{h}(i,j)$ indicates the weight score of the $i$-th query corresponding to the $j$-th key.
We then define a sparse mask matrix $SM^h$ of $h$-th head as follows:
\begin{equation}
{SM}^{h}_{(i,j)}= 
\begin{cases}
 1& \text{if } \mathcal{A}_{h}(i,j) \geq \mu_i \\ 
 0& \text{if } \mathcal{A}_{h}(i,j) < \mu_i
\end{cases},
\end{equation}
\begin{equation}
    \mu_i=\frac{1}{N}\sum_{j=1}^{N}\mathcal{A}_{h}(i,j),
\end{equation}
where $N$ is the length of the input sequence $\mathbf{x}$. Applied to $\mathcal{A}_{h}$, the $SM^h$ mask prunes its weak connections, whose weights are below the average attention weights $\mu_i$ along the key axis. 

During the experiment, we observed that using a common sparse mask combined along the head-axis outperforms applying $SM^h$ to each head individually.
In detail, we consider each head's activated positions for all the other heads; 
we define an adjusted sparse mask $SM_{OR}:=\bigcup_{h=1}^{H}SM^h$ where $H$ is the number of heads, and identically apply it to all attention heads. The $SM_{OR}$ mask is used during both training and inference.

\vspace{-2pt}
\subsubsection{Differentiable Pruning}
\vspace{-1pt}
\label{dp}
In the vanilla pruning (VP) method, the threshold of $SM^h$ is passively determined as the mean value of attention weights $\mu_i$.
However, the optimal threshold values vary depending on the number of layers, type of generation tasks, and degree of domain mismatch; thus, flexibly setting the threshold is preferable.
To this end, we propose a novel differentiable pruning (DP) method with learnable thresholds, inspired by \cite{Kim2022LearnedTP} which shares the same motivation in natural language processing task.

\begin{figure}[t!]
\vspace*{-16pt}
    \centerline{\includegraphics[width=.46\textwidth]{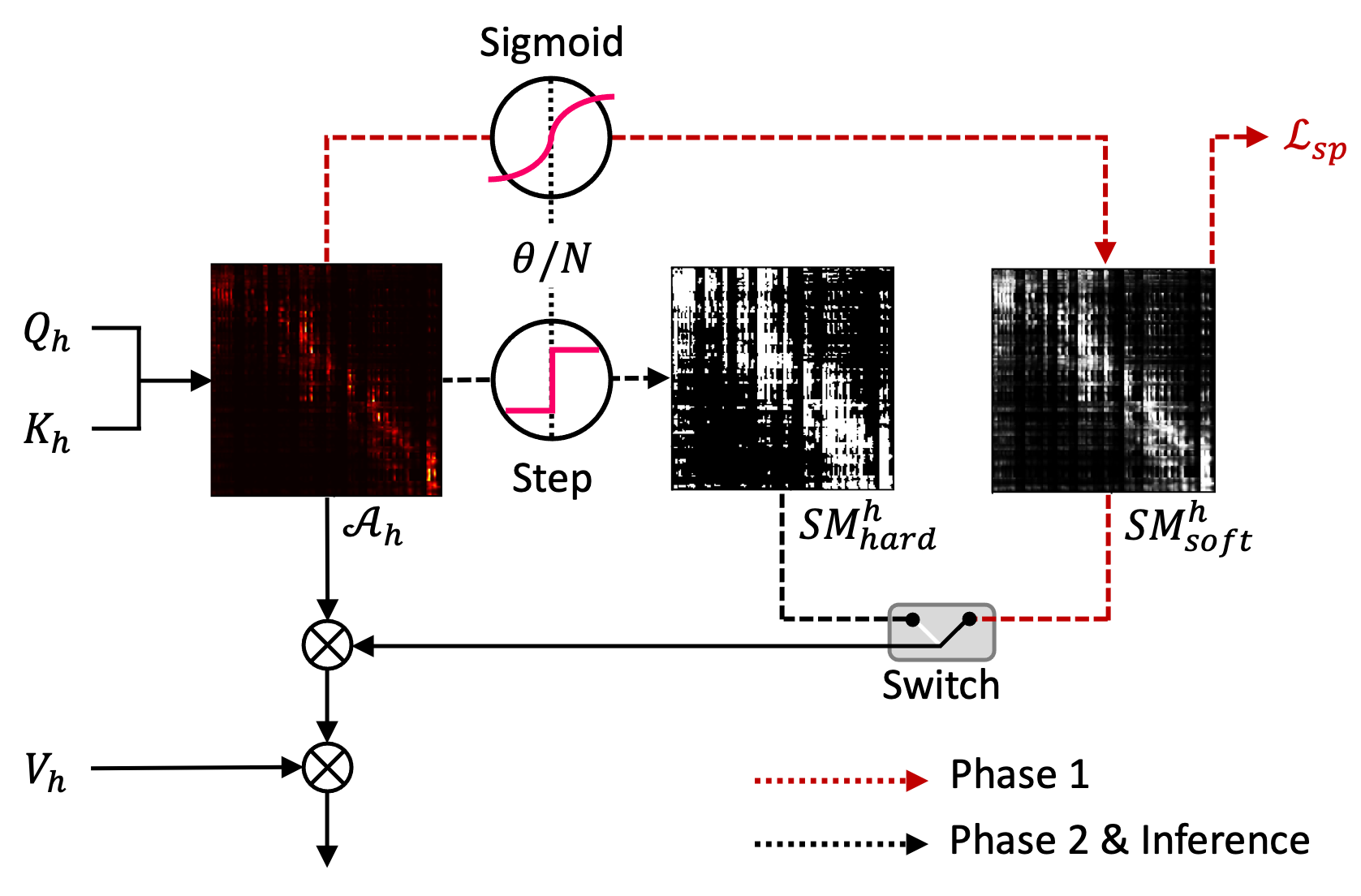}}
    \caption{\normalfont{Overview of the differentiable pruning. The soft sparse mask is used during phase 1 of training, and the hard sparse mask is used during phase 2 of training and during inference. The pruned attention head is multiplied with $V_h$ (values of $h$-th head) for succeeding process of self-attention operation.}}
    \label{fig:differentiable}
\vspace{-8pt}
    
\end{figure}

Fig. \ref{fig:differentiable} illustrates the overview of DP. 
In contrast to VP, which uses the predefined threshold, we first define a hard sparse mask of $h$-th head $SM^h_{hard}$ that inherits the learnable threshold $\theta/N$:
\begin{equation}
{{SM}^{h}_{hard}}_{(i,j)}= 
\begin{cases}
 1& \text{if } \mathcal{A}_{h}(i,j) \geq \theta/N \\ 
 0& \text{if } \mathcal{A}_{h}(i,j) < \theta/N
\end{cases},
\end{equation}
where $\theta$ is a trainable threshold parameter, and $N$ is the length of the sequence used to adjust the threshold value based on variations in input length. However, because the process of obtaining the binary mask $SM^h_{hard}$ is not differentiable, we cannot directly update $\theta$ by gradient descent. To solve this problem, we additionally adopt a differentiable soft sparse mask $SM^h_{soft}$ defined by a sigmoid function as follows:
\begin{equation}
    SM^h_{soft}=\sigma \left(\frac{\mathcal{A}_{h}-\theta/N}{T} \right),
\end{equation}
where $T$ is the temperature set to $0.01$ to approximate $SM^h_{soft}$ to $SM^h_{hard}$. The value of $SM^h_{soft}$ is close to $1$ where the attention weight is higher than the threshold $\theta/N$ and is close to $0$ in the opposite case.

We then propose a two-phase training method, summarized in Algorithm \ref{alg:differentiable}.
In phase 1, the entire model is trained with original TTS loss terms $\mathcal{L}_{tts}$ (i.e., loss terms of StyleSpeech\cite{min2021meta}) using the soft sparse masks $SM_{soft}$ to update model parameters including thresholds $\theta$.
We also add sparsity loss $\mathcal{L}_{sp}$ as a regularization term to ensure pruning behavior, as shown below:
\begin{equation}
\mathcal{L}_{sp}=\frac{1}{LH}\sum_{l=1}^{L}\sum_{h=1}^{H}\left({\overline{SM}}^h_{soft}-R\right)^2,
\end{equation}
where the sparsity ratio $R$ is a hyperparameter that indirectly determines the pruning strength; $L$ denotes the number of transformer layers, and $H$ represents the number of heads in each layer. Sparsity loss $\mathcal{L}_{sp}$ is defined as the average $L2$-distance between the soft sparse mask's mean value ${\overline{SM}}^h_{soft}$ and $R$ across all attention heads and decoder layers.
This loss term forces the model to be generalized to OOD data; without it, thresholds $\theta$ do not converge to meaningful values during training. This is because if only $\mathcal{L}_{tts}$ is used, the model obtains the lowest training loss value when \emph{no connections are pruned} (i.e., $\theta$ is stuck in $0$) for \emph{in-domain} data.
The value of $R$ is set between 0 and 1, and a lower R value prunes more connections. 
In summary, two loss terms are used when updating $\theta$: original TTS loss terms $\mathcal{L}_{tts}$ and the regularization term $\mathcal{L}_{sp}$. As mentioned previously, $\mathcal{L}_{tts}$ pulls $\theta$ down towards $0$, while $\mathcal{L}_{sp}$ aims to prevent this for generalization to OOD data. The threshold is consequently balanced by two opposing losses, pruning off self-attention connections only to the extent that it does not significantly harm the original objective of minimizing $\mathcal{L}_{tts}$. Thus, the degree of generalization can be controlled by varying the sparsity ratio $R$, while minimizing overgeneralization.

In phase 2, model parameters except $\theta$ are updated using the hard sparse masks $SM_{hard}$, whose thresholds $\theta$ are learned in phase 1. Here, $\mathcal{L}_{sp}$ is not used and the model is trained under the \emph{hard} pruning condition (low-weight connections are completely masked) with fixed pruning strength. 
This final model is used during inference.

\begin{algorithm}[t]
\caption{Differentiable pruning - Training procedure}\label{alg:differentiable}
\textbf{Phase 1}: Apply the soft sparse masks $SM_{soft}$, and update the model parameters including the thresholds $\theta$ with the original TTS loss terms $\mathcal{L}_{tts}$ and additional regularization term $\mathcal{L}_{sp}$

\textbf{Phase 2}: Apply the hard sparse masks $SM_{hard}$ determined by the learned thresholds $\theta$, and update the residual model parameters with training loss terms except $\mathcal{L}_{sp}$
\end{algorithm}

\if
\fi


%% file: 4_Experiment.tex
\section{Experiments}
\label{Experiments}

\subsection{Experimental Setup}
\textbf{Dataset and Preprocessing.} We used two subset datasets from LibriTTS\cite{zen2019libritts} (\textit{train-clean-100} and \textit{train-clean-360}) to train our model, which contain 245 hours of speech from 1151 speakers. For inference, VCTK\cite{yamagishi2019cstr} (108 unseen speakers) dataset is used for zero-shot TTS. 
For the method of preprocessing text and speech, we followed StyleSpeech \cite{min2021meta}.

\noindent\textbf{Model Details.}
We experimentally evaluated the performance of VP and DP considering StyleSpeech\cite{min2021meta} as a baseline. Consistent with \cite{min2021meta}, the encoder and decoder comprise 4 FFT blocks\cite{ren2020fastspeech} with 2 self-attention heads each.
For DP, 4 threshold parameters $\theta$ were declared in each decoder FFT block and were identically initialized to $0$. 
The training configurations for all implemented models were set to be the same as in \cite{min2021meta}, except that the models were trained for 300k steps.
In the case of DP, we advanced to phase 2 after training for 40k steps in phase 1.
For evaluation, we used the HiFi-GAN V1 vocoder\cite{kong2020hifi} to convert mel-spectrograms to audios.
In addition, two references were used for comparison: 1) ground truth audios and 2) audios generated by HiFi-GAN V1 (Voc.) conditioned on ground truth mel-spectrograms.

\noindent\textbf{Evaluation Metrics.}
Regarding subjective metrics, mean opinion score (MOS) evaluates the naturalness of speech, and similarity MOS (SMOS) evaluates speaker similarity.
Both metrics were scored on a scale of 1-5 by 16 raters, and we present them with 95\% confidence intervals (CI).
We used character error rate (CER) and speaker embedding cosine similarity (SECS) to evaluate intelligibility and speaker similarity as objective metrics. For CER, we transcribed the synthesized speech using the pre-trained speech recognition model provided by the SpeechBrain toolkit\cite{ravanelli2021speechbrain}. SECS is defined as the cosine similarity between speaker embeddings derived from the pre-trained speaker verification model\cite{ECAPA-TDNN} from \cite{ravanelli2021speechbrain}. Thus, MOS and CER assess speech quality, whereas SMOS and SECS assess similarity to the target speaker.
\vspace{-2pt}

\subsection{Evaluation on Zero-Shot TTS}
\vspace{-2pt}
For zero-shot TTS, we used arbitrary text input and randomly sampled one reference speech from each VCTK speaker for the reference encoder's input. 15 synthesized samples were used for MOS and SMOS, and 100 samples were used for CER and SECS.

From Table \ref{table:result}, we make the following observations: 1) The model with VP outperforms the baseline in all metrics except CER, demonstrating the generalization ability of the pruning method.
2) All models with DP remarkably surpass the baseline and the model with VP, particularly in terms of voice quality.
3) The results among models with DP show the trade-off relationship between pruning strength and performance. In the first viewpoint, the model is successfully generalized by pruning more connections ($R:0.50\rightarrow0.45$), resulting in a sharp increase in naturalness (+0.23 MOS). In contrast, excessive pruning ($R:0.45\rightarrow0.40$) rather reduces the model's original modeling capacity (i.e., overgeneralization); it causes a slight degradation in overall performance in our experiment. Intuitively, pruning all connections is the same as removing the entire self-attention module.

In summary, we conclude that DP significantly improves zero-shot TTS performance. Owing to its ability to adjust pruning strength, the model is also scalable to different degrees of domain mismatch (e.g., small $R$ in large domain mismatch). 
\begin{table}[t!]
\setlength\tabcolsep{2.8pt}
\begin{center} 
\caption{\normalfont{Comparisons of MOS, SMOS with 95\% CI, CER, and SECS results of zero-shot TTS; we used the StyleSpeech framework as the baseline system.
Note that VP and DP denote the vanilla and differentiable pruning techniques, respectively. 
For DP, we conducted 3 experiments by varying the sparsity ratio $R$. The best performances are in boldface.}}\label{table:result}
\begin{spacing}{1}
\begingroup
\begin{tabular}{l|cccc}
\toprule
\textbf{Model} & \textbf{MOS}($\uparrow$)  & \textbf{SMOS}($\uparrow$)  & \textbf{CER}($\downarrow$) & \textbf{SECS}($\uparrow$)   \\      
\midrule

Ground Truth     & 4.76$\pm$0.07 & - & - & -  \\
GT mel + Voc. & 4.67$\pm$0.08 & - & - & -  \\ 
\midrule
Baseline          & 3.43$\pm$0.12 & 2.99$\pm$0.16 & 4.56 & 0.268\\
VP                & 3.46$\pm$0.12 & 3.10$\pm$0.15 & 5.17 & 0.275\\ 
DP($R=0.50$)      & 3.53$\pm$0.12 & 3.18$\pm$0.15 & 3.96 & \textbf{0.279}\\ 
DP($R=0.45$)      & \textbf{3.76}$\pm$\textbf{0.11} & \textbf{3.23}$\pm$\textbf{0.15} & 3.96 & 0.278\\ 
DP($R=0.40$)      & 3.75$\pm$0.12 & 3.20$\pm$0.16 & \textbf{3.73} & 0.276\\ 
\bottomrule
\end{tabular}
\endgroup
\end{spacing}
\end{center}
\vspace{-14pt}
\end{table}

\begin{table}
\setlength\tabcolsep{3pt}
\begin{center}
\caption{\normalfont{MOS, SMOS with 95\% CI, CER, and SECS results of ablation studies. The best performances are in boldface.}}\label{table:ablation}
\begin{spacing}{1}
\begingroup
\begin{tabular}{l|cccc}
\toprule
\textbf{Model} & \textbf{MOS}($\uparrow$)  & \textbf{SMOS}($\uparrow$)  & \textbf{CER}($\downarrow$) & \textbf{SECS}($\uparrow$)  \\
\midrule
DP($R=0.45$)& \textbf{3.76$\pm$0.11} & \textbf{3.23$\pm$0.15} & \textbf{3.96} & \textbf{0.278}  \\
\midrule
w/o  ${SM}_{hard}$ & 3.65$\pm$0.11 & 3.02$\pm$0.16 & 4.21 & 0.274  \\
w/o  $\mathcal{L}_{sp}$ & 3.46$\pm$0.12 & 2.87$\pm$0.15 & 5.77 & 0.263 \\
\bottomrule
\end{tabular}
\endgroup
\end{spacing}
\end{center}
\vspace{-26pt}
\end{table}

\subsection{Ablation Study}
\vspace{-2pt}
Table \ref{table:ablation} shows the results of the ablation studies related to the two DP design techniques. We chose DP with $R = 0.45$ as the baseline because it performs best in terms of naturalness and similarity.
In the first experiment, we skipped the training phase 2 that uses the hard masks $SM_{hard}$; we only used the soft masks $SM_{soft}$ for training and inference. Results show that the two-phase training method is effective. Concretely, in phase 2, \emph{hard} pruning with updated thresholds improves the model's generalization performance by completely excluding low-weight connections during the text-to-mel conversion process.
In the second experiment, we removed the regularization term $\mathcal{L}_{sp}$, originally used in the training phase 1. Without $\mathcal{L}_{sp}$, the model shows poor performance because pruning does not occur at all. We also discovered that the thresholds $\theta$ were not updated from their initial value of $0$, as noted in section \ref{dp}. 
\vspace{-2pt}

\subsection{Analysis of Differentiable Pruning}
\vspace{-2pt}
To further analyze DP, we present the updated final thresholds $\theta$ of models with DP in Table \ref{table:visualization}. As expected, a smaller $R$ value generally leads to higher threshold values, indicating that more connections are pruned. Fig. \ref{fig:attn} represents the pruned attention heads of theses models using a specific text utterance and random reference speech.
The previously mentioned relationship between $R$ and pruning strength is also confirmed in the figure.
Remarkably, the pruned TTS models use only a few self-attention connections for high synthesis quality, implying that DP prevents the decoder from overfitting to in-domain data and improves the generalization performance. More materials of visualizations are in our demo page. 
\vspace{-1pt}

%% file: 5_Conclusion.tex
\section{Conclusion}
\label{Conclusion}
\begin{table}[t!]
\setlength\tabcolsep{3pt}
\begin{center}
\caption{\normalfont{Final DP thresholds $\theta$ updated in training phase 1.}}\label{table:visualization}
\begin{spacing}{0.9}
\begingroup
\begin{tabular}{lcccc}
\toprule
\multicolumn{1}{l}{\multirow{2}{*}{\textbf{Model}}} & \multicolumn{4}{c}{\textbf{Threshold} $\theta$}\\
\multicolumn{1}{c}{}                       & Layer \#1      & Layer \#2   & Layer \#3   & Layer \#4 \\\midrule
DP($R=0.50$) & 0.76 & 2.34 & 2.36 & 2.36  \\
DP($R=0.45$) & 1.70 & 3.53 & 4.18 & 4.18 \\
DP($R=0.40$) & 2.89 & 3.05 & 5.11 & 5.11 \\
\bottomrule
\end{tabular}
\endgroup
\end{spacing}
\end{center}
\vspace{-20pt}
\end{table}

\begin{figure}[t!]
\subfloat[$R=0.40$]{\includegraphics[width=0.183\textwidth]{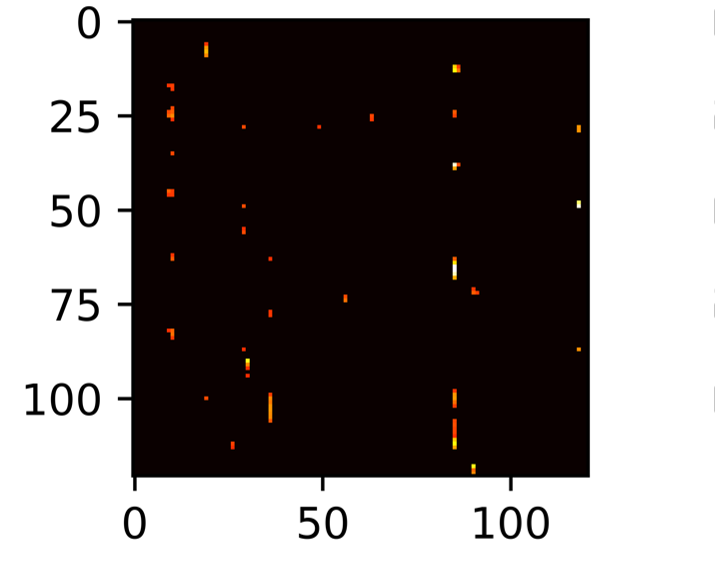}}
\hspace*{-15pt}  
\subfloat[$R=0.45$]{\includegraphics[width=0.183\textwidth]{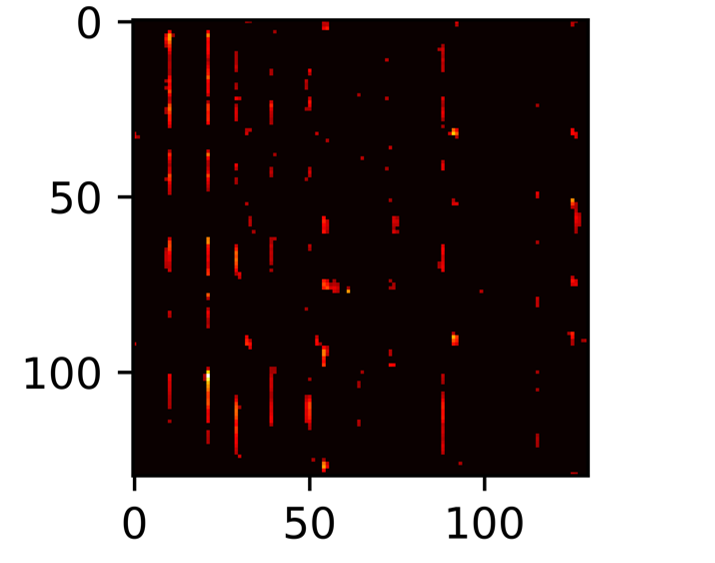}}
\hspace*{-15pt}  
\subfloat[$R=0.50$]{\includegraphics[width=0.183\textwidth]{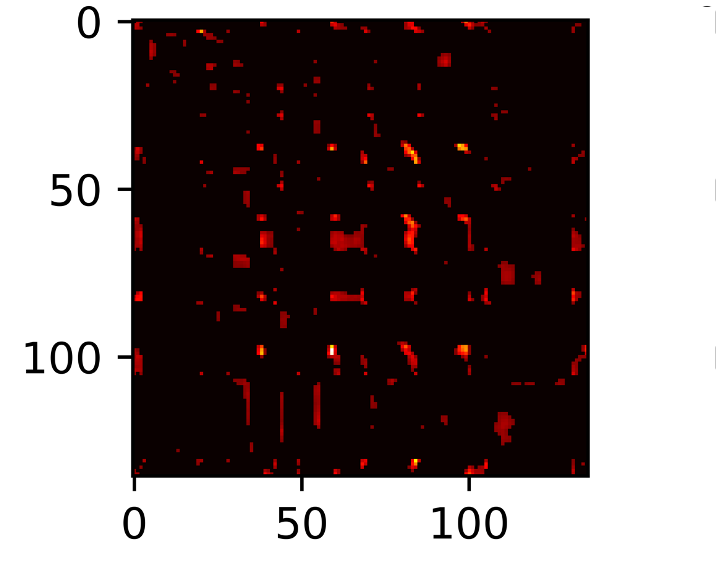}}
\caption{Pruned attention heads for the utterance “How many attention connections are pruned?”. Samples are first heads of the fourth decoder layers and are generated from DP with (a) $R=0.40$, (b) $R=0.45$, and (c) $R=0.50$. } \label{fig:attn}
\vspace{-15pt}
\end{figure}
\vspace{-2pt}
In this work, we proposed a self-attention pruning method for improving the generalization abilities of zero-shot multi-speaker TTS models. Furthermore, we investigated the optimal pruning techniques and emphasized the importance of differentiable pruning (DP), that can control the pruning strength augmented with the proposed two-phase training method. 
We then used it to generalize the mel-spectrogram decoder; evaluation on zero-shot multi-speaker TTS confirmed its superiority in terms of voice quality and speaker similarity. 
Future works include the application of DP for more severe domain mismatch cases.

%% file: 6_Ack.tex
\section{Acknowledgement}
\vspace{-2pt}
This work was supported by Voice\&Avatar, NAVER Cloud, Seongnam, Korea.